\documentstyle[preprint,aps]{revtex}
 \begin{document}
 \title{SAHA IONIZATION FORMULA AND THE VOIDS}
 \author{ MOFAZZAL AZAM}
 \address{THEORETICAL PHYSICS DIVISION\\
 BHABHA ATOMIC RESEARCH CENTRE,CENTRAL COMPLEX\\
 TROMBAY, MUMBAI-400085, INDIA}
 \maketitle
 \vskip .8 in

 \begin{abstract}
 The ultra-low density limit of Saha ionization formula
 suggests that, in this limit, matter would prefer to remain
 ionized.This has a very important implication for cosmic
 structures known as Voids.These are ultra-low density
 (much less than average density of matter in the Universe) regions
 in the galactic clusters and superclusters.The ionization
 formula implies that matter trapped in the Voids should be
 ionized.Therefore, we expect a very faint radiation glow
 from the Voids resulting from the motion of the charged
 particles.
 \end{abstract}
 \newpage
\section{Introduction}
 The Saha ionization formula \cite{saha20} has played a very impotant
 role in the development of astrophysics.The ultra-low density limit
 of this formula has been known for a long time \cite{feynman63}.In
 this limit, the formula suggests that the atoms in equillibrium
 prefer to remain  in ionized state.This ionization, just from
 "expansion" as the density goes down, has been listed as one
 of the surprises in theoretical physics by Peierls \cite{peierls79}.
 In this paper we point that this
 ultra-low density limit of Saha ionization formula is very relevant for
 the cosmic structures known as Voids.These ultra low density
 (much less than average density of matter in the Universe) regions
 dominate the volume in the Universe.
 The ionization
 formula implies that matter trapped in the Voids should be
 ionized.Therefore, we expect a very faint radiation glow
 from the Voids resulting from the motion of the charged

\section{Ultra-low desity limit of the Saha ionization formula}
 The ionization formula is given by,
 \begin{eqnarray}
 \frac{n_{e}n_{i}}{n_a}=\frac{1}{v_a}e^{-W/kT}
 \end{eqnarray}
 In the equation above, $n_e$ , $n_i$, $n_a$ are the densities of
 electrons, ions and atoms(not ionized) respectively. $W$ is the
 ionization potential, $T$ is the temparature and $k$ is the Boltzman
 constant.The volume occupied by a bound electron at
 temparature T is represented by $v_a$.
 It is, essentially, the
 volume contained within a thermal de Brogle wave lenghth.
 \begin{eqnarray}
 v_a=\lambda_{th}~^{3}=\Big(~\frac{2\pi \hbar^2}{m_{e}kT}~\Big)^{3/2}
 \end{eqnarray}
 Let us consider a box of volume $V$ which ,to start with,
 contains N number of hydrogen atoms.Let a fraction $X$ of them
 be ionized.In this case , $~n_e~=~\frac{N}{V}~X~=~n_i~$ and
 $~n_a~=~(1-X)~\frac{N}{V}~$.Substituting these values
 in the ionization formula we obtain,
 \begin{eqnarray}
 \frac{X^2}{1-X}~\frac{N}{V}~=~\frac{1}{v_a}e^{-W/kT}
\end{eqnarray}
 From the equation above, we see that the fraction of
 charged particles in equillibrium increases when we increase
 the volume(i.e., decrease the density).In the ultra-low density
 ($~\frac{N}{V}~ \rightarrow~~ 0$ or
 $~\frac{V}{N}~ \rightarrow~~\infty$ ),
 atoms would prefer to
 remain ionized \cite{feynman63,peierls79,ghosh98}.Before we discuss
 the nature and consequences of the ultra-low density limt,
 let us introduce the cosmic structures known as Voids
 \cite{zeldovich82,varun95} which
 are, essentially, underdense (less than the average density
 of matter in the universe) regions in galatic clusters
 and supeclusters.
 This requires review of some aspects of standard
 cosmology and the theory of structure formation at the large scale.We
 review very briefly, in next section,
 the materials relevant for our discussion.

 \section{The Voids}

 It is very well established through observation that
 at distance scales of the order of 200-300 megapersec, the Universe is
 isotropic and homogenous.This means that if we pick up a region of
 the Universe of dimension 200-300 megapersec at any distance and in any
 direction, it will contain the same amount of matter.Therefore, at this
 scale the density of matter can be considered to be constant.In Newtonian
 gravity, such a distribution
 of matter implies that at every point in space
 the potential and force are unbounded \cite{landau75}.
 This dilemma is resolved in the General
 Theory of Relativity.For an isotropic and homogenous
 distribution of matter one assumes the
 Friedman-Robertson-Walker metric, given by the line element
 \begin{eqnarray}
 ds^{2}= dt^{2}-a^{2}(t)(\frac{dr^{2}}{1-kr^{2}}+r^{2}d\theta^{2}+r^{2}\sin^{2} \theta d\phi^{2})
 \end{eqnarray}
 in which the Einstein equation,
 \begin{eqnarray}
 R_{\mu \nu}-\frac{1}{2} g_{\mu \nu}R=\frac{8\pi G}{3} T_{\mu \nu}
 \end{eqnarray}
 takes the simple  form \cite{landau75},
 \begin{eqnarray}
 \frac{\dot{a}^{2} +ka^{2}}{a^{4}} = \frac{8\pi G}{3} \rho_{0}
 \end{eqnarray}
 where $a(t)$ is the scale factor, $\rho_{0}$ is the averaged constant
 density, and $k=1, -1$ or $0$, respectively for closed, open and  flat
 universe.
 This equation along with the equation of state describes
 the isotropic and homogenous universe.The equations clearly show
 that the isotropic and homogenous
 distribution of matter can not be stable- the Universe
 expands.The constant density serves as a source term for the
 evolution of the scale factor.It does not give rise to attractive
 gravitational force.Therefore, the question that arises is:
 what is the source of gravity
 in the large scale?The answer is obtained as follows.
 When the mean free path of the particles
 is small, matter can be treated as an ideal fluid and the Newton's
 equations governing the motion of gravitating collisionless particles
 in an expanding Universe can be written in terms of
 $ ~{\bf x}= {\bf r}/a ~$ (the comoving space coordinate),
 ${\bf v} ={\bf \dot{r}}-H{\bf r}=a{\bf \dot{x}}$ (the peculiar velocity
 field, H is the Hubble constant),
 $\phi({\bf x},t)$ (the Newton gravitational potential) and
 $\rho({\bf x},t)$ (the matter density). This give us the following
 set of equations \cite{varun95,strauss95}.
 Firstly, the Euler equation,
 \begin{eqnarray}
 \frac{\partial (a {\bf v})}{\partial t}+({\bf v.\nabla_{x}}){\bf v}=
 -\frac{1}{\rho}{\bf \nabla_{x}} P-{\bf \nabla_{x}}\phi
 \end{eqnarray}
 Next the continuity equation
 \begin{eqnarray}
 \frac{\partial \rho}{\partial t}+ 3H\rho +\frac{1}{a}{\bf \nabla_{x}}
 (\rho {\bf v}) =0
 \end{eqnarray}
 And, finally the Poisson equation
 \begin{eqnarray}
 {\bf \nabla_{x}}^{2} \phi =4\pi Ga^{2}(\rho-\rho_{0})
 =4\pi Ga^{2} \rho_{0} \delta
 \end{eqnarray}
 where $\rho_{0}$ is the mean background density and
 $\delta=\rho/\rho_{0}-1$ is the density contrast.

 Therefore, at large scale, the source of gravity is not
 the average density $\rho_{0}$ but
 the density fluctuations, $\delta \rho >0$.
 It is a subject of study in theory of structure formation as to what
 kind of density fluctuation would grow in time and lead to the formation of
 galaxies, and clusters and superclusters of galaxies
 \cite{Padmanabhan93,Peebles80,varun95,strauss95}.
 It is important
 to remember that at the scale of dimensions, 200 - 300 megapersecs,
 the Universe
 is homogenous and isotropic and acquires constant density, and  therefore,
 if in some subregion $\delta \rho >0$, there must be some subregion where
 $\delta \rho <0$, so as to
 reproduce the constant density profile.These domains
 with $\delta \rho <0$ are known as {\bf Voids} \cite{zeldovich82,varun95}.
 Note that for Voids
 $ ~\delta\rho/\rho_{0} ~$ is always bounded bellow
 by $-1$ .Such regions of Voids
 dominate the volume in the universe giving rise to cellular structures
 with the clusters and superclusters of galaxies forming string like walls
 around them.Existence of Voids are supported by direct observation as well
 as numerical simulation of hydrodynamic equations
 \cite{varun95,ryden96,hoyle01,antonu00}. The observed Voids seem to have
 dimension of several (tens of) megapersecs.

 \section{Conclusion}

 The Voids are the ultra-low density regions in the Universe,
 and these are the regions where one would expect to observe
 the consequences of ultra-low density limit of the Saha ionization
 formula.As discussed before, in this limit, atoms would prefer to
 remain ionized.At this stage, the important question is: what is the
 source of the ionization energy? There are the starlights but at high
 red shift, their intensity is very low.
 The most common source of ionization
 energy at high red shift is the lights from Quasars.
 \par The second source
 which may sound somewhat speculative is the following.In the begining
 the Voids expand faster than the Universe \cite{varun95}.However, in
 the radiation domination era the whole Universe is kept in a
 single equillibrium state by the radiation field.During the decoupling
 of radiation, this equillibrium is destroyed, and in the process
 some ions may remain trapped in the Voids.
 \par The Saha ionization  formula implies that , at ultra- low density,
 once the ionization takes place there is hardly any chance for
 recombination \cite{feynman63,peierls79,ghosh98}.Therefore, the motion
 of the charged particles in the Voids should create a faint
 radiation glow
 
\end{document}